\documentstyle[multicol,psfig,aps,prb]{revtex}

\newcommand{\ba}{\begin{eqnarray}}                                             
\newcommand{\ea}{\end{eqnarray}}

\newcommand{\dg}{$^{\circ}$ }
\newcommand{\dge}{$^{\circ}$}

\newcommand{\be}{\begin{equation}}
\newcommand{\ee}{\end{equation}}
\newcommand{\la}{\langle}

\newcommand{\ra}{\rangle}

\newcommand{\mgm}{MgB$_2$ }
\newcommand{\mge}{MgB$_2$}

\begin{document}
\draft

\title{An angle-resolved soft x-ray spectroscopy study of the
electronic states of single crystal \mge}

\author{G. P. Zhang,$^{1,2,*}$ G. S. Chang,$^1$ T. A. Callcott$^1$}

\address{$^1$Department of Physics and Astronomy, The University of
Tennessee at Knoxville, TN 37996} 

\address{$^2$Department of Physics, Indiana State University, Terre
Haute, IN 47809$^*$}

\author{D. L. Ederer}
\address{Physics Department, Tulane University, New Orleans, LA 70118}

\author{W. N. Kang, Eun-Mi Choi, Hyeong-Jin Kim, Sung-Ik Lee}

\address{National Creative Research Initiative Center for
        Superconductivity and Department of Physics, Pohang University
        of Science and Technology, Pohang 790-784, Korea}

\date{\today}

\maketitle

\begin{abstract}

Angle-resolved soft x-ray measurements made at the boron K-edge in
single crystal \mgm provide new insights into the B-2p local partial
density of both unoccupied and occupied band states. The strong
variation of absorption with incident angle of exciting x-rays permits
the clear separation of contributions from $\sigma$ states in the
boron plane and $\pi$ states normal to the plane. A careful comparison
with theory accurately determines the energy of selected critical $k$
points in the conduction band. Resonant inelastic x-ray emission at an
incident angle of 15\dg shows a large enhancement of the emission
spectra within about 0.5 eV of the Fermi level that is absent at 45\dg
and is much reduced at 60\dge.  We conclude that momentum transferred
from the resonant inelastic x-ray scattering (RIXS) process couples
empty and filled states across the Fermi level.

\end{abstract}


\begin{multicols}{2}

\section{Introduction}

The discovery of superconductivity in \mgm came as a
surprise.\cite{mgb2} Its unusually high transition-temperature
immediately motivated intensive investigations.  Although the
conventional BCS mechanism most likely underlies this peculiar
superconductivity, the precise mechanisms are less clear. Different
sample qualities and different experimental techniques show
inconsistent gap values ranging from 2 meV to 7 meV.\cite{gap}
However, in a single crystal sample, results are less ambiguous.  A
recent angle-resolved photoemission (PE) study on a small \mgm single
crystal showed detailed dispersion of the occupied band along the
$\Gamma-K$ and $\Gamma-M$ directions.\cite{pe} However, due to the
large background and low intensity for normal emission, the dispersion
along the $k_z$-direction was not resolved.

Soft x-ray absorption (SXA) spectroscopy \cite{xray} complements PE by
providing an element and angular momentum selected local, partial
density of states (LPDOS) of the unoccupied bands.  Similarly, soft
x-ray fluorescence (SXF) spectra provide a measure of the LPDOS for
the occupied valence bands.  Although these spectroscopies are not
generally selective for momentum, they can often provide an accurate
measure of density of states features associated with critical points
of the band structure and thus provide a detailed check on theoretical
band calculations. These features of x-ray spectroscopies are clearly
illustrated in a recent paper (the first paper of Ref. [5]) on
\mge,\cite{prb_mgb2} which used SXF and SXA studies to map the boron
p-LPDOS for a polycrystalline sample.

We have recently shown that in an anisotropic material such as
NaV$_2$O$_5$,\cite{prl_nav2o5} additional selectivity can be obtained
by varying the incident angle of the exciting
x-rays. \cite{prb_nav2o5} By varying the incident angle of exciting
x-rays so that the polarization vector is varied with respect to a
bond direction, it is possible to selectively excite electrons to
electronic orbitals having particular spatial orientation. In this
paper, we report angle resolved soft x-ray absorption and emission
measurements at the boron $K$-edge of a single crystal film of
\mge. The soft x-ray absorption spectra provide a measure of the B-2p
LPDOS for unoccupied states above the Fermi level. The angular
resolved spectra allow us to separate excitation to band states
derived from B $\sigma$ states directed within the boron plane and
$\pi$ states directed perpendicular to the plane. The results enable
us to identify critical points of unoccupied band states along several
crystal-momentum directions, which can be directly compared to our
theoretical calculations.

For excitation near a band edge, the excitation and emission processes
may be coupled so that the emission spectra must be interpreted in
terms of a resonant inelastic x-ray scattering (RIXS) process.  In the
RIXS process incident x-rays excite electronic excitations and are
scattered with reduced energy.  Previous RIXS experiments in
graphite,\cite{graphite} diamond, silicon, cBN\cite{cbn} and
hBN\cite{hbn} show that in well screened and delocalized electronic
systems the momentum conservation inherent in the RIXS process can
provide direct information about band dispersion.  These analyses
generally assume that the momentum transferred in the RIXS process is
small on the scale of the Brillouin zone so that the electronic
excitations induced by RIXS are nearly vertical in the reduced
zone. In \mge, however, we observe a strong enhancement of a narrow
band of states near the top of the occupied band that requires an
electronic excitation across the Fermi level and thus a transfer of
momentum.  The variation of this enhancement with incident angle
demonstrates that the enhancement is a result of momentum transfer
from the RIXS process, and not primarily associated with other
processes such as phonon-assisted transitions that might allow
momentum to be conserved.

The paper is arranged as follows. In Section II, we present the
angular-resolved absorption spectra. The angular-resolved RIXS spectra
are presented in Section III. 

\section{Angular resolved x-ray absorption spectra}

A single crystal of \mgm was grown as a thin film of 400 nm thickness
on the $R$-plane ($1\bar{1}02$) of an Al$_2$O$_3$
substrate.\cite{sample} An amorphous B thin film was deposited by
pulsed laser deposition and a high-quality and single crystal \mgm
film was obtained by sintering the thin film at 900$^{\circ}$C for 20
minutes in Mg vapor. The crystal quality and orientation was confirmed
by x-ray diffraction.  The process is described in detail
elsewhere.\cite{sample} The dc magnetization measured with a
superconducting quantum interference device (SQUID) magnetometer
showed that the superconducting transition temperature (Tc) is 39 K,
and the 10-90\% transition width is 0.7 K.  Our soft x-ray
measurements were made at room temperature at Beamline 8.0 of the
Advanced Light Source at Lawrence Berkeley National Laboratory.  The
slits of the beamline monochromator were set to provide a resolution
of about 0.1 eV for the absorption spectra, while the resolution of
the emission spectrometer is about 0.2 eV at the 190 eV energy of the
Boron K spectra.

The experimental geometry is indicated in Fig. 1.  \mgm is a layered
compound with alternating layers of boron and Mg stacked along the $c$
axis. Graphite-like hexagonal layers of boron lie in the $a-b$
plane. Linearly polarized x-ray light from an undulator and
monochromator was incident on the sample. By rotating the sample
holder, we can scan the polarization vector from the $a-b$ plane
toward the $c$ axis with $\theta$ changed from 15\dg to 45\dge.
$\theta $ measures the angle between the incident x-ray beam and the
sample normal as well as the angle between the polarization vector of
incident photons and the $a-b$ plane.

Figure 2(a) shows the dramatic change of the absorption spectra
measured in total fluorescent yield mode with $\theta$.  The spectra
shown are taken at 7.5\dg intervals from 15\dg to 45\dg with angles
measured from the $a-b$ plane.  The experimental spectra of Fig. 2(a)
are normalized with respect to the total areas from 183.55 to 208.55
eV (boron $K$-edge).  Experimental and theoretical spectra are aligned
with peak labeled $a$.  The Fermi level is set to zero (long dashed
line).  The zero energy corresponds to an energy of 187.28 eV with
uncertainty of about 1 eV in the absolute energy, consistent with our
previous measurement.\cite{prb_mgb2} The relative energies of emission
and absorption spectra are calibrated to be accurate to about 0.1 eV.
Figure 2 shows that the Fermi level cuts through the pre-edge of peak
$a$, a result which agrees with the photoemission data \cite{pe} and
shows normal metallic behavior. We identify six spectral features
labeled from $a$ to $f$.  At $\theta=15$\dge, only two peaks $a$ and
$f$ are observable leaving a 4 eV-wide valley from 1 eV to 5 eV.  As
the polarization rotates from the $a-b$ plane toward the $c$ axis,
peak $a$ grows slightly while peak $f$ drops. The most remarkable
change is in the valley where the whole valley is gradually filled. We
will concentrate on the two broad peaks at 2.2 and 2.8 eV above the
Fermi level, which increase in amplitude to a point comparable to peak
$a$.  We note here that in a polycrystalline or powder sample these
angular dependent changes in the absorption spectra are not observed
because they are intrinsically angle averaged.\cite{refereea}

Our calculated {\it ab initio} results are shown in Fig. 2(b), where
 we also present two additional spectra at $\theta=0$\dg and 90\dge.
 For an easy comparison, we plot spectra calculated for the exact same
 experimental conditions.  The angular variation results from explicit
 calculation of the dipole matrix elements taking into account the
 projection of the polarization vector on the orbital amplitude in
 real space. The theoretical calculations have been described in
 greater detail in a recent paper on NaV$_2$O$_5$.\cite{prb_nav2o5} In
 brief, we calculate the eigenvalues and wavefunctions by solving the
 Kohn-Sham equation, \be \hat{H}\Psi_{n{\bf k}}({\bf r})=\{
 -\frac{{\hbar}^2}{2m}{\nabla}^2+ \hat{V}_{\rm eff}\} \Psi_{n{\bf
 k}}({\bf r})=E_{n{\bf k}}\Psi_{n{\bf k}}({\bf r})~. \nonumber \ee
 Then, we compute transition matrix elements between the core level
 and unoccupied bands.  It is quite impressive that the theoretical
 spectra reproduce the experimental ones very accurately. Both the
 overall shape and peaks and valleys are consistent with the
 experimental counterparts.  A quantitative comparison is given in
 Table I. One notices that except for peak $f$, theoretical results
 for peaks and valleys from $b$ to $e$ are about 0.2 eV larger than
 the experimental ones, which may suggest band narrowing possibly due
 to correlation effects.  This band narrowing, if real, is much
 smaller than the simple metals such as Na, Li and Mg,\cite{na}
 especially considering the experimental uncertainty of about 0.2 eV.
 In any case, the effect is small.

\subsection{Incident at $\theta=15$\dge}

The good agreement between theory and experiment paves the way for a
detailed investigation of the origin of the peaks.  To help our
analysis, we draw several vertical short-dashed lines across both the
absorption spectra [Figs. 2(a) and 2(b)] and the band structure
[Fig. 2(c)] to highlight important connections. We label those
crossing points of peaks $a$, $c$, $d$ and $f$ by $a_1$, $a_2$,
etc. With an incident angle of 15\dge, the polarization vector is
15\dg from the $a-b$ plane, so that we mostly probe the
$\sigma$-orbitals directed between B atoms lying in the a-b plane.
For excitation to states within 0.5 eV of the absorption edge, we see
from the spectra that only peak $a$ is excited.  Now comparing with
the band structure, we can see six different points from $a_1$ to
$a_6$ which potentially contribute to this peak and distribute almost
evenly along all momentum directions. For instance, points $a_1$ and
$a_4$ in Fig. 2(c) fall near the $\Gamma$ point along the $\Gamma-M$
and $\Gamma-K$ lines respectively, while $a_2$ is close to the $M_1$
point. $a_3$ is almost in the middle of the $\Gamma-K$ line, and
$a_5$, representing two nearby points along the $A-L$ line, is close
to the $A$ point, whereas $a_6$ along the $A-L$ line is close to the
$L$ point.

If we solely depended on the energy position, we would not be able to
resolve these different band contributions to the $a$ peak, a common
complication in soft x-ray spectroscopy. Fortunately, \mgm is highly
anisotropic so that there are strong variations in the angular
dependent transition matrix elements that allow us to resolve the
contributions from different bands. We can use the transition matrix
elements $\la \psi_{nk}(r)|\vec{E}\cdot \vec{r}|\psi_{B1s}(r)\ra$ to
resolve these features, where $|\psi_{nk}(r)\ra$ is the band state and
$|\psi_{B1s}(r)\ra$ is the boron 1s core level. Transition matrix
elements contain both the direct space and reciprocal space information,
which helps us to probe the spatial orientation of orbitals and reveal
the dispersive band in momentum space.  

In Fig. 3, in the upper part, we show the BZ in the three dimension,
while in the lower part, we plot the calculated transition matrix
elements at various positions in the BZ and for different angles of
the incident light. The four columns represent the transition matrix
elements for four peaks from $a$ to $f$ in the momentum space. Rows 1
and 2 refer to $k_z=0$ plane at the bottom of the BZ which contains
the $\Gamma$ point while 3 and 4 refer to $k_z=2\pi/c$ plane at the
top of the BZ which contains the $A$ point (see the top graph in
Fig. 3).  The first and third rows refer to the results at
$\theta=15$\dge, while the second and fourth ones the results at
$\theta=45$\dge.  The high symmetry $k$ points are also indicated in
Figs. 3(a) and 3(c) for the top and bottom planes. Numbers near the
contours in Fig. 3 indicate the magnitude of transition matrix
elements (in arbitrary units).  First focus on Figs. 3(a) and 3(c),
which show the matrix elements contribution to peak $a$ for
$\theta=15$\dge. From those two figures, one can see clearly that the
largest transition matrix elements are well localized at the $\Gamma$
point and in the vicinity of the $A$ point while contributions from
other directions are very small.  In three dimensions, it is found
that the matrix elements remain equally strong for the region
surrounding the $\Gamma-A$ line that lies within the tubular Fermi
surface \cite{band} that surrounds this line (see the top graph of
Fig. 3).  Returning to Fig. 2(c), now we can identify that band states
at $a_1(a_4)$ and $a_5$ contribute most strongly to peak
$a$. Therefore, the position of peak $a$ can be used to measure the
position for $\Gamma_1$, which is 0.27 eV above the Fermi surface. Our
assignment is also consistent with the nature of the $\sigma$ band
along the $\Gamma_1-A_1$ line. It is well known that this famous
$\sigma$ band is derived from electronic orbitals in the $a-b$
plane. Since our present electric field polarization is nearly in the
$a-b$ plane, this band is strongly excited.

If we increase the incoming photon energy to 0.69 eV (with respect to
the Fermi level), the spectra start to fall sharply since there is no
density of states above $A_1$ and bands along $M-K-\Gamma$ can not be
effectively excited (see below). When we increase the photon energy
further up to the $\Gamma_2$ point, we begin to access a new band
($c_4$ and $c_5$, or $d_4$ and $d_5$), thus we would expect a strong
increase of the spectral weight. Surprisingly, we see a 4 eV-wide
valley instead. The valley also appears in the theoretical spectra
(see Fig. 2(b)).  The transition-matrix element contours in Figs. 3(e)
and 3(g) are for peak $c$, and in Figs. 3(i) and 3(k) for peak $d$,
but the matrix elements are small.

In order to obtain a qualitative understanding of these variations in
the matrix elements, we calculated the spatial charge density
distribution. For the bands at point $c_6$ (similarly $d_6$, $c_2$ and
$c_3$), we find that they are derived from $\pi$-orbitals and are
highly oriented along the $c$ axis (see Fig. 4(a)), which explains why
they can not be effectively excited. However, this does not explain
bands at points $c_4$, $c_5$, $d_4$ and $d_5$ since they do lie in the
$a-b$ plane (see Fig. 4(b)). A very careful investigation reveals the
secret: those orbitals in Fig. 4(b) are strongly localized around Mg
sites, not B sites. Since our measurement is element-specific and is
done at the boron $K$ edge, we are not sensitive to orbitals around
the Mg sites, which explains the appearance of the 4-eV wide valley at
$\theta=15$\dge.

Above the valley labeled $e$ at 3.66 eV, the spectra increase in
intensity.  Further increasing the photon energy to 6.0 eV above the
Fermi level, we pick up high-lying $\sigma$ bands along the
$\Gamma_3-A_2$ direction and the $\Gamma-K$ (A-L) line. This can be
seen directly from the transition-matrix contour plot in Figs. 3(m)
and 3(o). Therefore, the energy position of peak $f$ is a good measure
of the position of the $\Gamma_3$ and $A_2$ points. Our experiment
gives the value of 6.65 eV, which can be compared with the theoretical
value of 6.55 eV.

\subsection{Incident at $\theta=45$\dg}

When we rotate the polarization vector from the $a-b$ plane toward the
$c$ axis ($\theta=45$\dge), we begin to probe orbitals which are
oriented along the $c$ axis.  We start from peak $a$. Comparing
Figs. 3(a) with 3(b) (Figs. 3(c) with 3(d)), one notices that the
spectral weight moves away from the $\Gamma$ (A) point to the M-K
(H-L) line. Qualitatively we can say that the contribution to this
peak from the $\sigma$-bands is decreasing with angle while the
contribution from the $\pi$-bands bands is strongly increasing.
Clearly, these angular measurements give us the ability to cleanly
separate the contribution of bands with different spatial orientation
to the measured density of states.

As the angle is increased, two broad peaks $c$ and $d$ appear at 2.2
and 2.8 eV above the Fermi surface (see Figs. 2(a) and 2(b)),
respectively.  Although there are six crossing points contributing to
peak $c$, we can safely exclude $c_1$, $c_4$ and $c_5$ since they are
from Mg sites.  The transition matrix contour in Fig. 3(f) shows that
states with large elements are near the K point, while at
$k_z=2\pi/c$, the matrix elements are only half that of the states at
$k_z=0$ (see Fig. 3(h)).  Therefore, peak $c$ predominantly comes from
states at $c_2$ and $c_3$ and determines their energy positions.  The
momenta for these states are (0.38,0.25,0) and (0.3,0.3,0),
respectively.  They form a hot spot which leads to peak $c$.  For peak
$d$, the transition matrix contour plot (see Fig. 3(j)) shows that the
dominant contribution comes from states along the $M-K$ and $\Gamma-K$
lines.  States at $d_4$ and $d_5$ (see Fig. 2(c)) do not contribute
because those states originate from Mg sites. The main contribution is
from bands at $d_2$ and $d_3$ which are highly oriented along the $c$
axis (see Fig. 4(a)).  They form another hot spot which ultimately
leads to peak $d$. By carefully analyzing the data, we find that point
$d_3$ is at (0.28, 0.28, 0) in the reciprocal space while point $d_2$
is at (0.46, 0.09,0), very close to the $M$ point. The position of
peak $d$ of 2.8 eV is a good measure of the energy position for the
$M_2$ point.

Summarizing all the data, we tabulate in the last column of Table I
the measured values of unoccupied band states at selected points of
both high and low symmetry in k-space. These measured values of
particular points in k-space provide values that may be compared
directly to calculated values.

\section{Angle resolved resonant inelastic scattering}

We now turn our attention to the soft x-ray (SX) emission spectra
produced when electrons from occupied valence states refill the
excited core state.  Well above the excitation threshold, the
excitation and emission processes are always decoupled so that the
normal SX fluorescence (SXF) spectra represent the boron p-LPDOS of
the filled valence band.  Near threshold, however, the absorption and
emission processes are often strongly coupled and must be described in
terms of a resonant inelastic x-ray scattering (RIXS) process. An
important defining feature of RIXS spectra is that the energy
difference between incident and scattered electrons is equal to the
energy of the electronic excitations so that RIXS features move in
energy with the excitation energy, while normal SXF features remain
fixed in energy.  For near threshold excitation, normal SXF and RIXS
are often superimposed. In the present data, resonant features in the
SXF spectra are observed near threshold for excitation to the $\sigma$
states that contribute to peak $a$ in Fig. 2. For energies more than 1
eV above the absorption threshold, the spectra do not change with
further increases in energy and are representative of SXF spectra.

The emission spectra are presented in Fig. 5 for excitation energies
within 1 eV of the boron K edge. Note that the Fermi level is at about
187.28 eV.  Spectra are shown for incident angles of 15\dg and
60\dge. Figure 5(a) shows the results at 15\dg where we have seen that
excitation is mainly into the $\sigma$ states. When we excite at
187.25 eV, only an elastic peak A can be seen. Increasing the
excitation energy to the threshold of 187.50 eV, a shoulder appears on
the low energy edge of A and a broad peak B appears around 185.05 eV.
Its energy position stays the same even with higher excitation
energies of 188.00 and 188.25 eV, which identifies it as a normal
fluorescence peak, where the excitation and de-excitation processes
are decoupled.

From Fig. 5(a), we notice that at the excitation energy of 187.50 eV,
the elastic peak A becomes asymmetric and broader. The shape of its
left shoulder indicates a possible peak hidden there.  This is indeed
the case.  When we further increase the excitation energy to 187.75
eV, peak C is revealed at 187.30 eV. Two significant features of this
peak may be noted. The strong resonant enhancement of the peak remains
visible so long as absorption is into the $\sigma$-band and ends when
the excitation energy exceeds the $A_1$ point of this band (see
Fig. 2). The resonance region extends only to about 1 eV below the
Fermi edge and thus does not include the lower energy $\sigma$ states.
With excitation at an incident angle of 60\dg shown in Fig 5(b), the
resonant enhancement is greatly reduced, but is still present. For
excitation at 45\dg (not shown), no enhancement at all is observed in
the SXF spectra. Comparing Figs. 5(a) with 5(b), we find that they are
very similar, and within our experimental resolution of 0.2 eV the
peak positions are identical (see Table II). The only significant
difference is the reduction in the magnitude of the resonant
enhancement near the Fermi level when the angle is increased from
15\dg to greater angles.

It is difficult to account for the resonant enhancement in terms of an
ordinary RIXS process in a delocalized system, because k-conservation
is expected for the two photon scattering process.  K-conservation
follows from the Kramers-Heisenberg formula that describes the RIXS
process \cite{prl_nav2o5} \ba S(\omega,\omega')\propto\sum_f\left |
\sum_m\frac{\la f| p\cdot A |m\ra\la m|p\cdot
A|gs\ra}{\omega+E_{gs}-E_m-i\Gamma}\right |^2 \nonumber\\
\times \delta(E_{gs}+\omega-E_f-\omega')~, \ea where $|gs\ra$, $|m\ra$, and
$|f\ra$ are initial, intermediate, and final states, and $E_{gs}$,
$E_m$, and $E_f$ are their energies, respectively; $\omega$ and
$\omega'$ are the incident and emitted photon energies; $\Gamma$ is
the spectral broadening due to the core lifetime in the intermediate
state. $p\cdot A$ is the transition operator.  If $|\psi_{nk_1}\ra$ is
the intermediate state $|m\ra$ with momentum $k_1$ and
$|\psi_{nk_2}\ra$ is the final state $|f\ra$ with momentum $k_2$,
besides the energy conservation, the total momentum should be
conserved, i.e., $q+k_1=q'+k_2$, where $q$ and $q'$ are momenta for
the incoming and outgoing photons, respectively.  If the photon
momenta $q$ and $q'$ were negligible as is the case in visible
spectroscopy, we would have $k_1\approx k_2$ or zero momentum
transfer.

Previous calculations\cite{band} show that the empty states from the
$\sigma$ band lie in a small tubular region in k-space surrounding the
$\Gamma$-A axis,\cite{band} while the filled states from this band are
located outside of this tube.  This behavior can also be seen for the
$\Gamma$-K, $\Gamma$-M and A-L axes in Fig. 2(c) and are similar for
all other directions normal to the $\Gamma$-A axis.  Thus the resonant
enhancement requires a mechanism that provides momentum normal to this
axis and couples the empty states inside the tube to the filled states
outside the tube.

First, we will estimate the magnitude of the energy loss and momentum
transfer for the enhanced emission. When excited at 187.75 eV, the
resonant inelastic peak C in Fig. 5(a) has the energy loss that is
centered at about 0.5 eV and covers a range of about 1 eV. To excite
this range of energies requires a momentum transfer in the range of
0.1 on the scale of 2$\pi/a$ which sets the scale of the BZ. All
subsequent momentum values are quoted on this scale.

We have considered two possible processes that might provide the
required momentum.  They are phonon assisted transitions and momentum
transfer from the RIXS scattering process.\cite{refereeb} A previous
calculation of the phonon spectrum\cite{phonon} shows that the
in-plane phonon mode $E_{2g}$ has a very strong electron-phonon
coupling (EPC) along the $\Gamma$-A, $\Gamma$-M, and A-L
directions. Such a strong but selective coupling might provide way to
transfer momentum from the phonon subsystem to the electronic
subsystem.  Since there is no valence state available along the
$\Gamma$-A direction, its strong coupling is irrelevant to the
anomalous enhancement.  The most important coupling is along the
$\Gamma$-M and A-L directions.  To be more specific, the in-plane
$E_{2g}$ phonon mode with the strongest EPC is the most
important\cite{phonon}. Along the $\Gamma$-M direction, phonon can
transfer momentum ${\bf q}$ to maximum values of (0.15,0,0), while
along the A-L direction, transfers occur to maximum values of ${\bf
q}=(0.2,0,0.5)$.  These momentum transfers are of the correct order of
magnitude to efficiently couple the excitation and de-excitation
processes.

Since momentum is transferred in the SX scattering process, we
also considered the possibility that this accounts for the observed
enhancement. As indicated in Fig. 6(a), the scattering of X-ray photons
through 90\dg requires a transfer of momentum in the interaction with
the surface along a line 45\dg from the incident direction.  For 15\dg
angle of incidence on the sample, the momentum transfer normal to the
$\Gamma$-A line is calculated to be about 0.03 (2$\pi/a$), while for a
45\dg angle of incidence, there is no momentum transfer normal to this
line. The mechanism is illustrated schematically in Fig. 6(b).

The variation of the magnitude of the resonant enhancement as the
incident angle is varied allows us to choose between these two
mechanisms.  A phonon-assisted process should not depend on angle so
that the enhancement in the spectra should only be reduced moderately
by the change in the projection of the polarization on the $\sigma$
orbitals in the $a-b$ plane. If the momentum is provided by the RIXS
scattering process, the enhancement should be absent at 45\dg and
greatly reduced at 60\dge.  The measured enhancement is indeed found to
be unobservable at 45\dg and greatly reduced at 60\dge. We conclude
that the enhancement is enabled by the momentum provided by the RIXS
process, allowing non-vertical transitions within the BZ.

We note finally that two separate angular dependent effects are
present in the RIXS spectra. First the projection of the polarization
vector of the incident x-rays on the orbitals determines the intensity
of the absorption into the derived intermediate states.  This factor
falls off as a function of $\cos\theta$ with increasing incident
angle.  And second, the resonant enhancement near threshold depends on
the momentum transfer perpendicular to the $\Gamma-A$ axis provided by
the RIXS process, which varies as $\cos(\theta-45^{\circ})$ with a
90\dg scattering angle.  A new experimental endstation will soon be
available that will permit the scattering angle in RIXS measurements
to be varied over a wide range, which will greatly facilitate the
study of momentum transfer effects.

We acknowledge many helpful discussions with Eric L. Shirley.  This
research is supported by NSF Grant No. DMR-9801804. Measurements were
carried out at the Advanced Light Source at LBNL supported by DOE
contract DE-A003-76SF00098.  This work is also supported in part by
the Ministry of Science and Technology of Korea through the Creative
Research Initiative Program.

$^*$Mailing address.

\end{multicols}

\begin{table}
\begin{caption}
{A comparison between experimental and theoretical energy positions
for six peaks and valleys. The last column shows their corresponding
crystal momenta and assigned special symmetry points. All the energies
are referenced to the Fermi level. }
\end{caption}

\vspace{1cm}

\begin{tabular}{ccccl}
Peak \& valley &~~~Experiment (eV)~~~&~~ Theory (eV)~~&~~Deviation (eV)~~&$k-$point\\
\hline
a& 0.27&           0.27&0.00&$\Gamma_1$ \\
b& 1.25  &         1.49&0.24&$\Gamma_2$\\
c& 2.24    &       2.43&0.19& (0.38, 0.25, 0) \\ 
 &         &           &    & (0.3, 0.3, 0)\\       
d& 2.83     &      2.96&0.13& (0.46, 0.09, 0)($\rm M_2$)\\
 &         &           &    & (0.28, 0.28, 0)\\       
e& 3.66      &     3.79&0.13& $\rm M_3$\\
f& 6.65       &    6.55&-0.10& $\Gamma$, ($\rm A_2$) \\
\end{tabular}

\vspace{2cm}

\begin{caption}
{ The resonant inelastic peak positions for two different angles
$\theta=15$\dg and $\theta=60$\dge.  At $\theta=15$\dg the strong
electron-phonon coupling in the flat $\sigma$ band leads to a surprising
enhancement of peak C at 187.30 eV. The energy units are in eV.  The
experimental uncertainty is about 0.2 eV.}
\end{caption}

\vspace{1cm}

\begin{tabular}{lllll}
Excitation energy  &\multicolumn{2}{c}
{$\theta=15$\dge} &\multicolumn{2}{c} {$\theta=60$\dge} \\
\hline
A & B  &   C             & B  &  C      \\
\hline
187.50  & 185.05  &                & 185.05  &        \\
187.75  & 185.05  & 187.30         & 185.05  & 187.07 \\
188.00  & 185.05  & 187.21         & 185.05  & 186.95 \\
188.25  & 185.05  & 186.98         & 185.05  & 186.95 
\end{tabular}

\end{table} 

\begin{figure}
\caption{Experimental geometry.  A linearly polarized photon with
energy $\omega_{in}$ is incident on the sample, and the outgoing
photon energy becomes $\omega_{out}$.  ${\bf n}$ is surface normal,
${\bf E}$ is polarization of incident photon, $(a-b)$ is plane of
boron atoms, and ${\bf c}$ is normal to boron plane.  }
\end{figure}

\begin{figure}
\caption{ (a) The angular-resolved experimental absorption spectra at
the boron $K-$edge. $\theta$ represents the angle between the electric
polarization $\rm \vec{E}$ and the $a-b$ plane. We scan from 15\dg to
45\dg with incremental angle of 7.5\dge. Six peaks and valleys are
labeled by $a-f$.  The experimental geometry is shown in the
inset. (b) Theoretical spectra with the exact same condition. Two
dashed curves represent results at 0 and 90\dge, respectively.  (c)
Band structure. Vertical short-dashed lines highlight the connection
between the band structure and absorption spectra. Crossing points are
denoted with $a_1$, $a_2$, $a_3$, etc. The long vertical line
represents the Fermi level which is set to zero. }
\end{figure}

\begin{figure}
\caption{ Transition-matrix element contour in the reciprocal
space. The $x$ axis denotes the $k_x$ while the $y$ axis the $k_y$
axis. Four columns represent results for peaks $a$, $c$, $d$ and $f$,
respectively. The first and third rows are calculated at
$\theta=15$\dge, while the second and fourth rows are at
$\theta=45$\dge. The first and second rows are for the crystal
momentum $k_z=0$ while the third and fourth rows are for
$k_z=2\pi/c$. Numbers (in arbitrary units) near contours show the
magnitude of transition matrix elements. The high symmetry points are
shown in (a) and (c), and are also shown in 3-dimensional BZ.  }
\end{figure}

\begin{figure}
\caption{ Unoccupied spatial charge density plot in the $y-z$ plane
(the $z$ axis is same as the $c$ axis).  (a) represents charge
distribution for bands at point $c_3$ (see Fig. 1(c)) and other points
$c_2$, $d_2$ and $d_3$. The charge is mainly located around boron
sites. (b) represents those bands at point $c_4$ (see Fig. 1(c)) as
well as other points $c_1$, $c_5$, $d_1$, $d_4$, and $d_5$. The
density is located around magnesium sites, which makes little
contributions to our absorption spectra. }
\end{figure}

\begin{figure}
\caption{ Resonant inelastic x-ray scattering spectra measured at (a)
$\theta=15$\dg and (b) $\theta=60$\dge. The main difference is that at
$\theta=15$\dge, there is an enhancement of the inelastic peak C at
excitation energy of 187.75 eV. The Fermi energy is at about 187.28
eV. }
\end{figure}

\begin{figure}
\caption{(a) Momentum transfer $\Delta \bf k$ in RIXS process.  The
projection of the momentum transfer along the a/b plane changes with
the incident angle, with the minimum at 45\dg in our experimental
geometry.  $\bf k_{in}$ and $\bf k_{out}$ are incident and outgoing
momenta, respectively.  ${\bf n}$ denotes the surface normal or the
$c$ axis.  (b) With momentum transfer $\Delta k$, valence electrons
from different bands can recombine with the core hole, showing a
strong enhancement in the RIXS spectra (see Fig. 5(a)).  }
\end{figure}

\newpage

\hspace{5cm}\psfig{figure=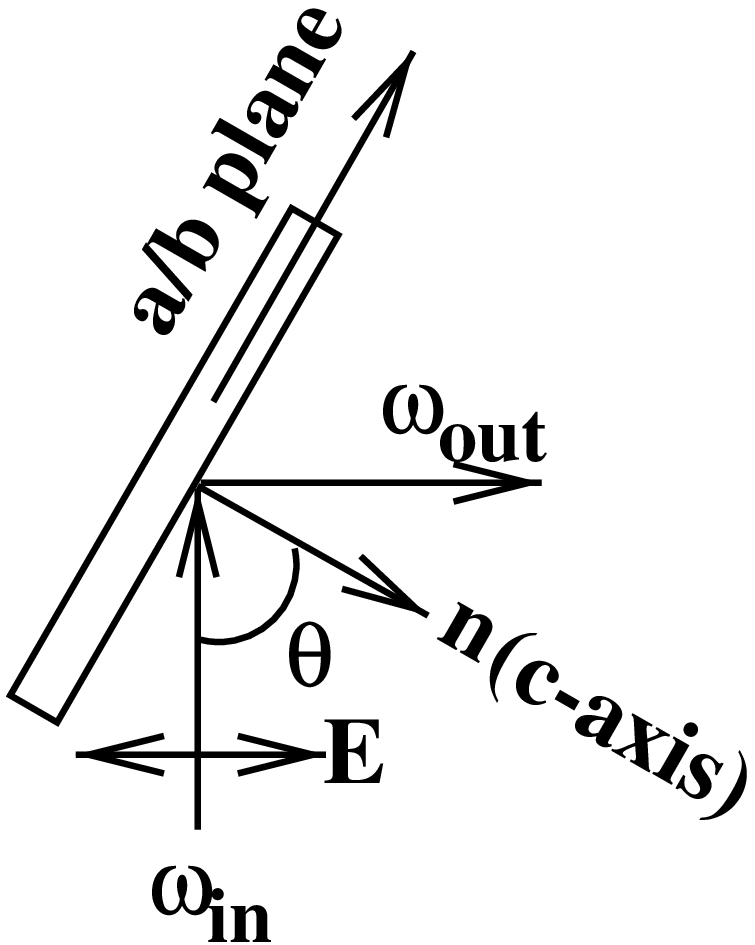,width=8cm,angle=0}

\vspace{4cm}

\centerline{FIGURE 1}

\hspace{1cm}\psfig{figure=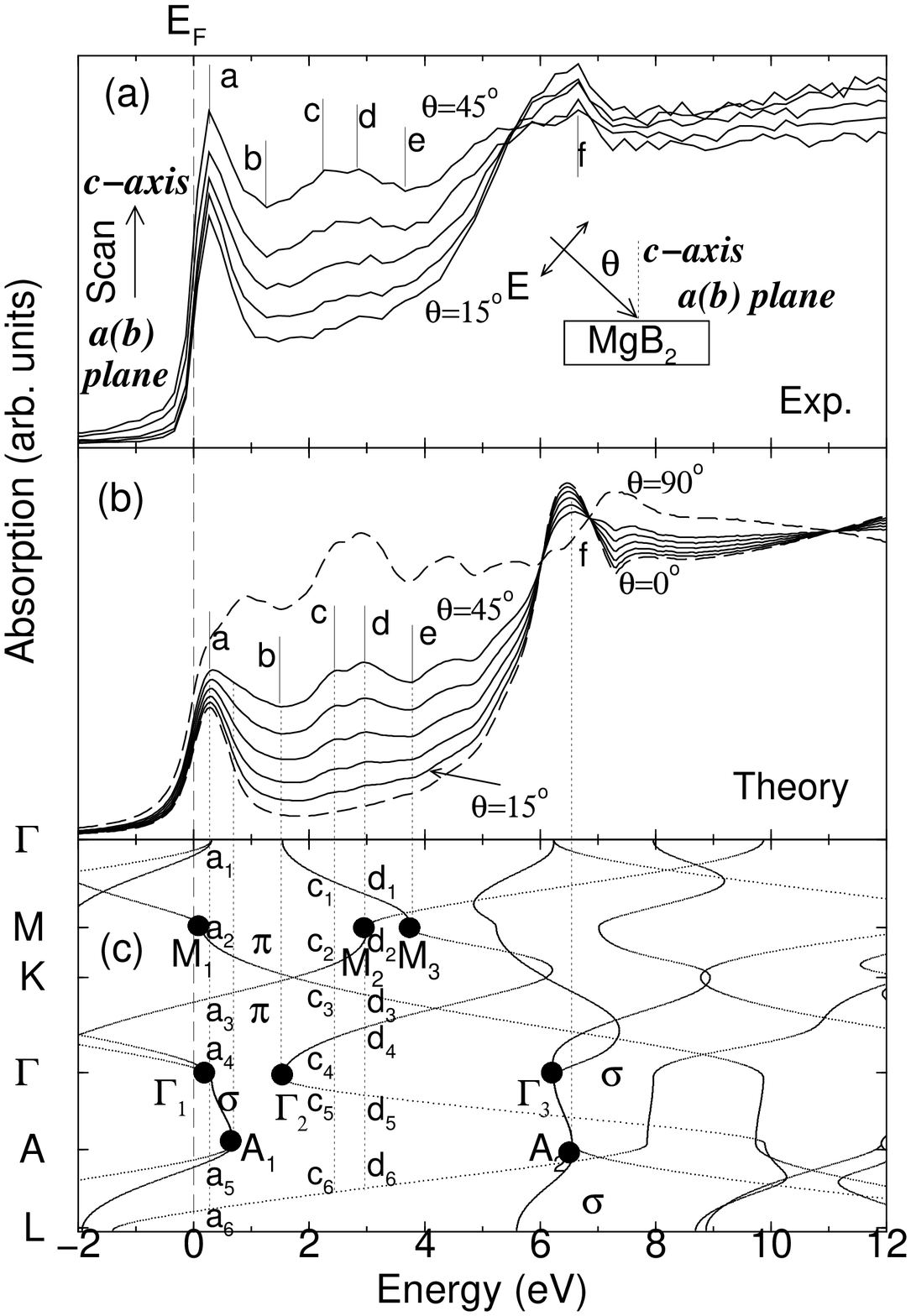,width=12cm,angle=180}

\vspace{1cm}

\centerline{FIGURE 2}

\newpage

\hspace{7.5cm}\psfig{figure=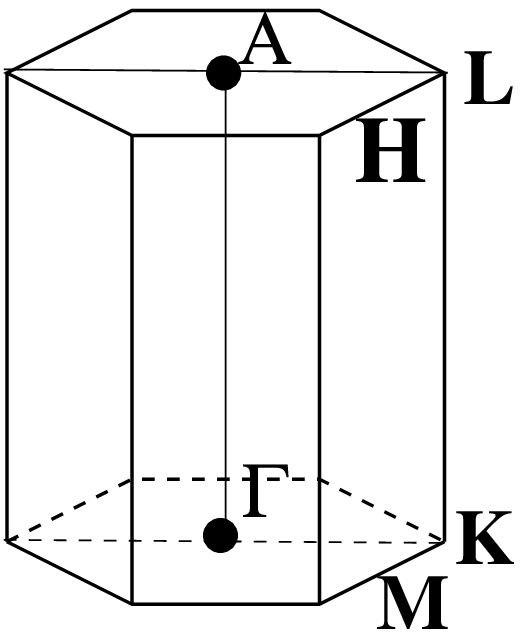,width=3cm,angle=0}

\vspace{-0.5cm}

\hspace{1.5cm}\psfig{figure=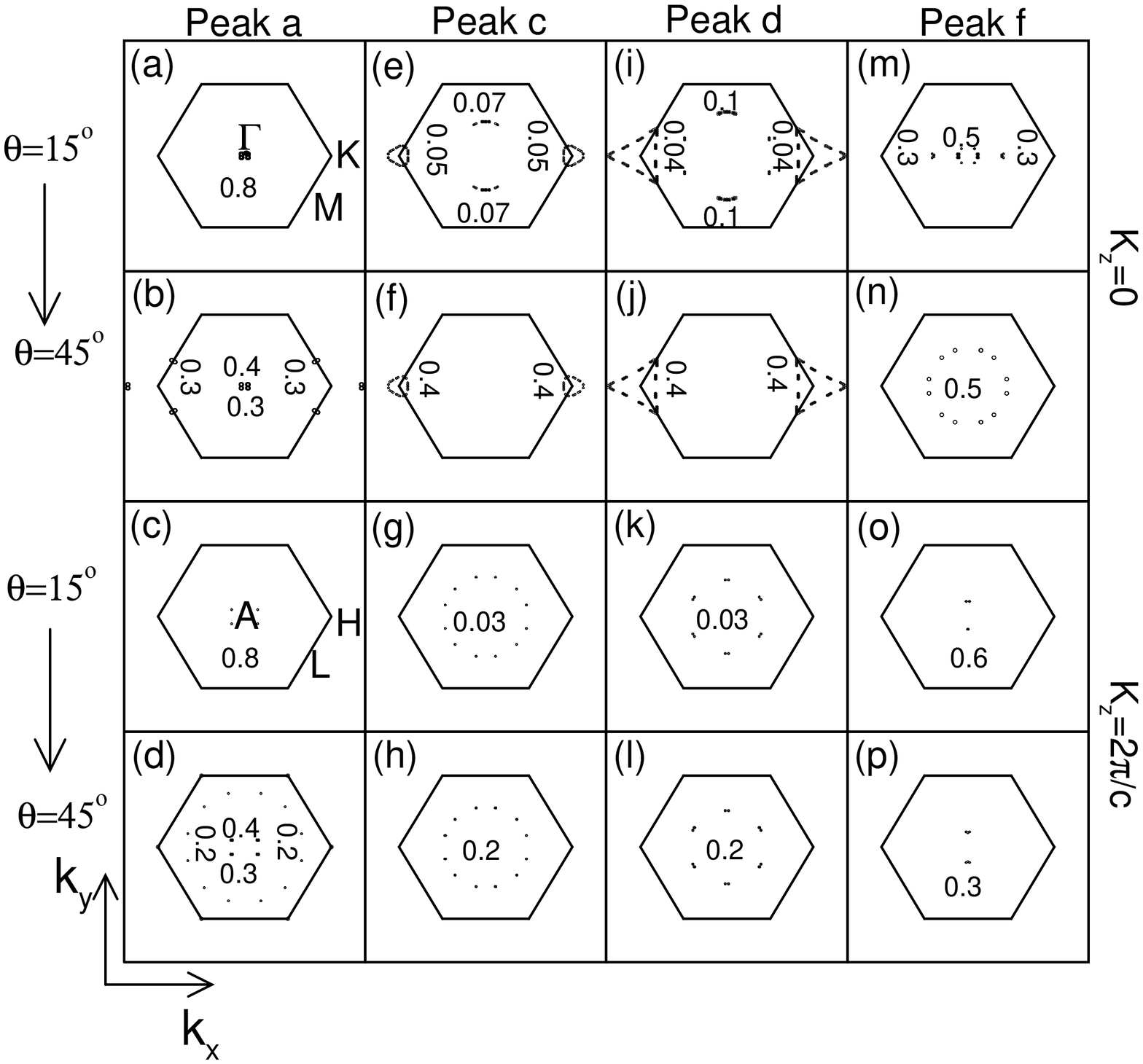,width=18cm,angle=270}

\vspace{1cm}

\centerline{FIGURE 3}

\hspace{0.8cm}\psfig{figure=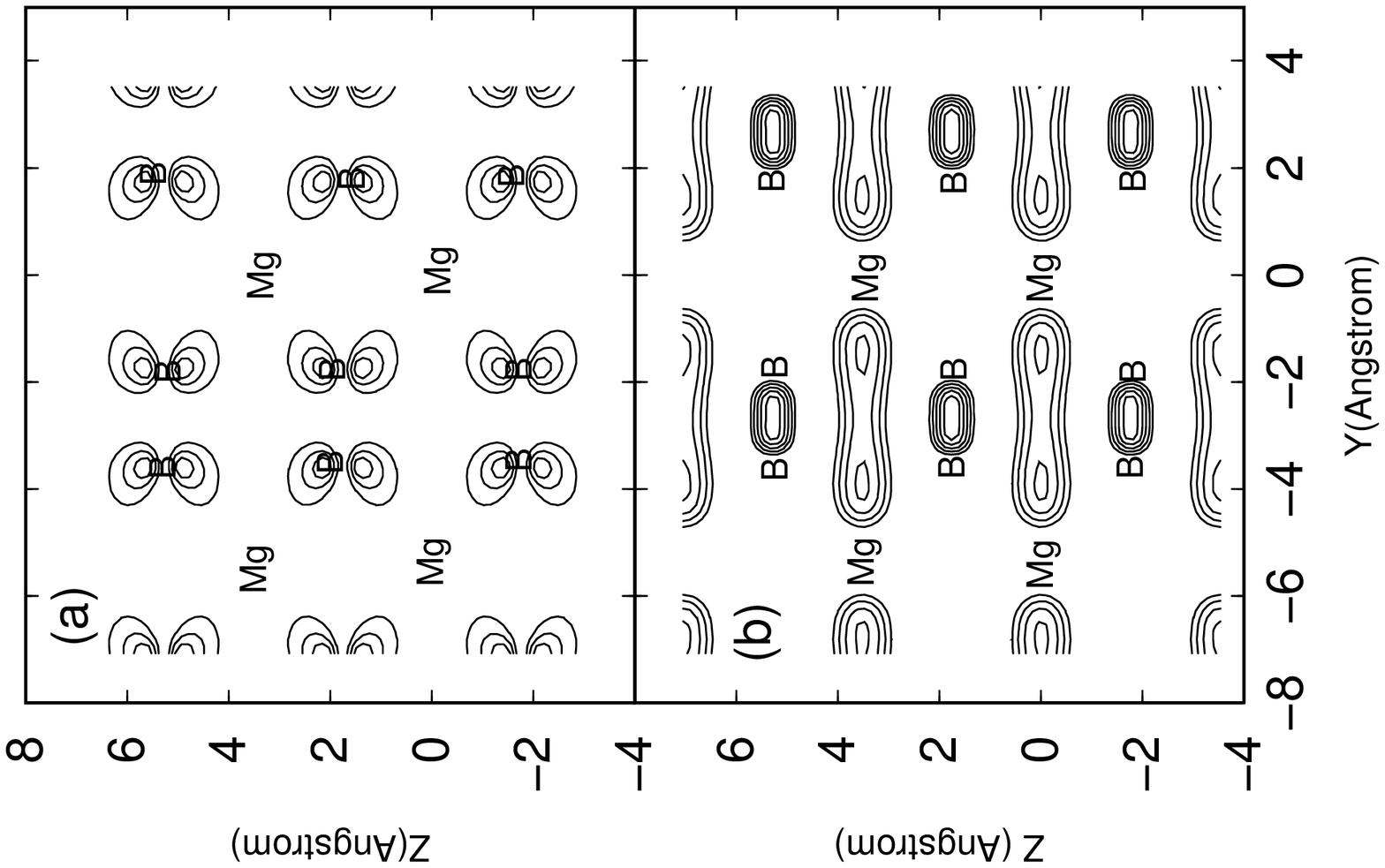,width=12cm,angle=270}

\vspace{2cm}

\centerline{FIGURE 4}

\hspace{-1.5cm}\psfig{figure=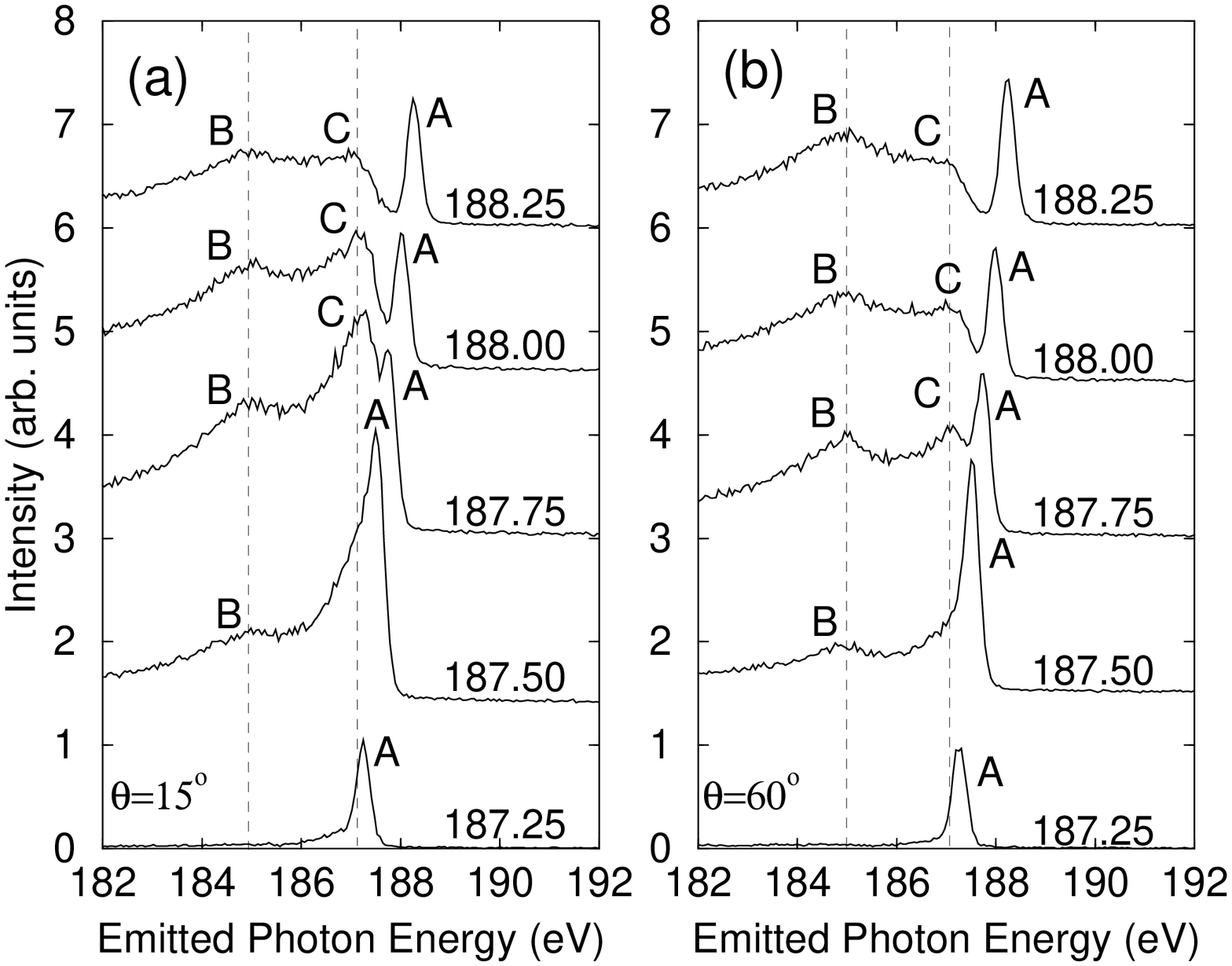,width=17cm,angle=270}

\vspace{2cm}

\centerline{FIGURE 5}

\hspace{-.5cm}\psfig{figure=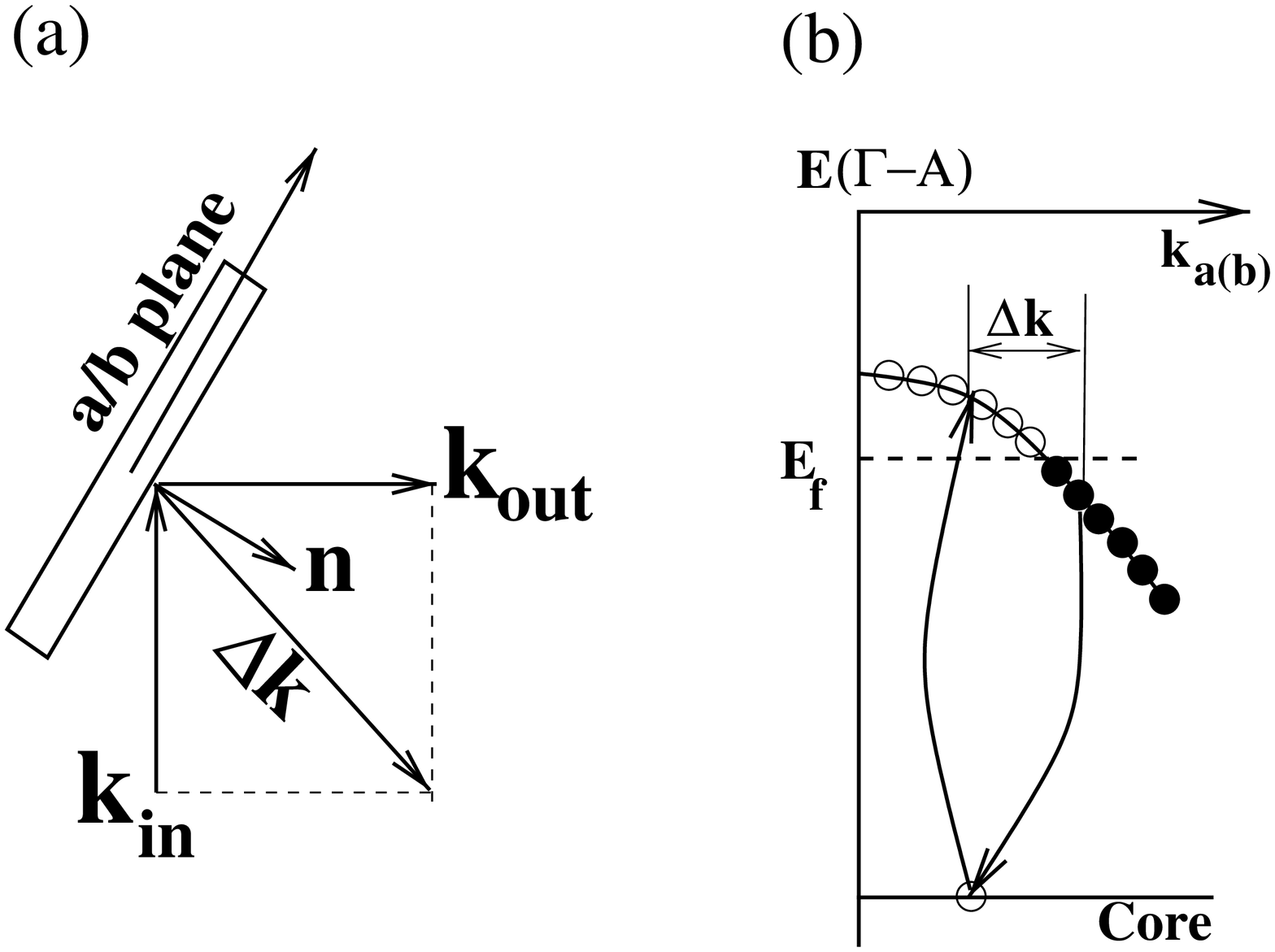,width=17cm,angle=0}

\vspace{2cm}

\centerline{FIGURE 6}


\begin{references}
\bibitem{refereeb} 
The self-absorption in our RIXS data is very weak.  The
reason is as follows.  First, the self-absorption effect becomes
stronger with increase in the incident angle as seen from the
absorption spectra. Consequently, if the self-absorption had a strong
contribution to our RIXS data, we would expect a monotonic decrease in
the intensity of peak C. This does not agree with our experimental
observation: the intensity of peak C decreases and after 45 degrees,
it increases.  Therefore, we conclude that the self-absorption effect
less likely plays a dominant role in our RIXS spectra.
\end{references}
\end{document}